\documentclass[prl,twocolumn,amsmath,amssymb,floatfix]{revtex4-1}
\usepackage{graphicx}
\usepackage{dcolumn}
\newcolumntype{d}[1]{D{.}{.}{#1}}
\usepackage{longtable}

\usepackage{epstopdf}

\newcommand{\tns}{Ta$_2$NiSe$_5$}
\newcommand{\bog}{$B_{1g}$\ }
\newcommand{\btg}{$B_{2g}$\ }
\newcommand{\brg}{$B_{3g}$\ }
\newcommand{\bou}{$B_{1u}$\ }
\newcommand{\btu}{$B_{2u}$\ }
\newcommand{\bru}{$B_{3u}$\ }
\newcommand{\ag}{$A_{g}$\ }
\newcommand{\bg}{$B_{g}$\ }
\newcommand{\au}{$A_{u}$\ }
\newcommand{\bu}{$B_{u}$\ }

\begin{document}


\title{Orthorhombic-to-monoclinic transition in Ta$_2$NiSe$_5$ due to
  a zone-center optical phonon instability}

\author{Alaska Subedi} 

\affiliation{CPHT, CNRS, Ecole Polytechnique, IP Paris, F-91128
  Palaiseau, France} 
\affiliation{Coll\`ege de France, 11 place
  Marcelin Berthelot, 75005 Paris, France}

\date{\today}

\begin{abstract}

I study dynamical instabilities in Ta$_2$NiSe$_5$ using density
functional theory based calculations.  The calculated phonon
dispersions show two unstable optical branches.  All the acoustic
branches are stable, which shows that an elastic instability is not
the primary cause of the experimentally observed
orthorhombic-to-monoclinic structural transition in this material.
The largest instability of the optical branches occurs at the zone
center, consistent with the experimental observation that the size of
the unit cell does not multiply across the phase transition.  The
unstable modes have the irreps $B_{1g}$ and $B_{2g}$.  Full structural
relaxations minimizing both the forces and stresses find that the
monoclinic $C2/c$ structure corresponding to the $B_{2g}$ instability
has the lowest energy.  Electronic structure calculations show that
this low-symmetry structure has a sizable band gap.  This suggest that
a $B_{2g}$ zone-center optical phonon instability is the primary cause
of the phase transition.  An observation of a softening of a $B_{2g}$
zone-center phonon mode as the transition is approached from above
would confirm the mechanism proposed here.  If none of the $B_{2g}$
modes present in the material soften, this would imply that the
transition is caused by electronic or elastic instability.

\end{abstract}


\maketitle

\section{Introduction}

There have been continual experimental efforts at finding excitonic
insulators, which are electron-hole condensate analogues of
superconductors, since they were theoretically proposed to exist in
the 1960s in small-band-gap semiconductors or semimetals
\cite{kozl65,jero67}.  \tns\ shows a second-order phase transition
with a resistivity anomaly near $T_c =$ 328 K \cite{suns85,disa86},
and Wakisaka \textit{et al.}\ have proposed that this material is an
excitonic insulator based on their observation of an extremely flat
valence band edge below $T_c$ \cite{waki09}.
%
%
Additional experimental features purporting to support an excitonic
insulator phase in \tns\ have been reported
\cite{seki14,kim16,lu17,mor17,lark17,werd18a,mor18,li18,werd18b,seo18, lark18,
  okaz18,fuku19,lee19}.
%
%
However, collective phenomena like Meissner effect and
dissipationless flow that are observed in superconductors and
superfluids have not yet been observed in \tns.
%

There seem to be no symmetry arguments forbidding the formation of a
complex order parameter of the type $\Delta e^{i\theta}$ in excitonic
insulators.  However, there are experimental indications that such a
complex order parameter cannot explain the low-temperature phase of
\tns.  The high- and low-temperature phases of \tns\ occur in the
$Cmcm$ orthorhombic and $C2/c$ monoclinic structures, respectively
\cite{suns85,disa86}.  The low-temperature phase arises when the $m_x$
$(x \rightarrow -x)$ and $m_z$ $(z \rightarrow -z)$ mirror symmetries
of the high-temperature phase are broken, and a deviation from
90$^\circ$ of the angle $\beta$ between the axes $a$ and $c$ that
breaks those symmetries was observed in the 1980s
\cite{suns85,disa86}.  This occurs when an elastic strain
$\varepsilon_{B_{2g}} \equiv \varepsilon_{xz}$ with the irreducible
representation (irrep) \btg spontaneously develops in the
high-temperature orthorhombic structure as its $c_{55}$ elastic
modulus softens and becomes unstable while the $T_c$ is crossed.
Nakano \textit{et al.}  have observed such a softening of the
transverse acoustic mode corresponding to the $c_{55}$ elastic modulus
in their inelastic x-ray scattering measurments \cite{naka18a}.  In
addition, they have also determined intra-unit-cell antiferroelectric
atomic displacements that break the mirror symmetries.  If the
$\varepsilon_{B_{2g}}$ strain is not the primary order parameter, the
softening of the $c_{55}$ elastic modulus can only occur when there is
a linear coupling between the order parameter and the strain
\cite{rehw73}.  However, a complex order parameter cannot couple
linearly to strain and, hence, cannot be the cause of the spontaneous
development of the $\varepsilon_{B_{2g}}$ strain in \tns.  This has
previously been pointed out by Zenker \textit{et al.}\ \cite{zenk14}.
Nevertheless, several theoretical studies have proposed a complex
order parameter for the low-temperature phase
\cite{kane13,ejim14,sugi16a,yama16,domo16,sugi16b,mats16,domo18,sugi18}.

Nakano \textit{et al.}\ find that the antiferroelectric displacement
pattern observed in the low-temperature phase could be decomposed to
two $B_{2g}$ phonon modes of the high-temperature phase that
respectively involve movements of Ta and Se ions \cite{naka18a}.
Although they observed a strong softening of the transverse acoustic
mode with the irrep $B_{2g}$ corresponding to the $c_{55}$ elastic
modulus as the $T_c$ is approached from 400 K, they did not observe a
similar softening in the only $B_{2g}$ optical phonon mode that they
could resolve between 30 and 100 cm$^{-1}$.  However, the optical
modes do exhibit a large linewidth broadening.  This suggests the
presence of a large electron-phonon coupling and indicates that an
electronic instability possibly drives the phase transition in \tns.

The fact that any $B_{2g}$ instability of the $Cmcm$ phase breaks the
$m_x$ and $m_z$ mirror symmetries has important implications on the
electronic structure of \tns.  The highest-lying valence and
lowest-lying conduction bands belong to different irreps in the
high-temperature $Cmcm$ phase \cite{kane13}.  Therefore, they cannot
hybridize to open a gap. When the $m_x$ and $m_z$ mirror symmetries
are broken, the two bands can hybridize and lead to a gap opening.
Mazza \textit{et al.}\ \cite{mazz19} and Watson \textit{et
  al.}\ \cite{wats19} have both highlighted the importance of the loss
of the two mirror symmetries in describing the low-symmetry $C2/c$
phase.  Mazza \textit{et al.}\ constructed a minimal model based on Ta
and Ni $d_{xz}$ orbitals and studied the effects of on-site $U$ and
nearest-neighbor $V$ Coulomb interactions using variational
Hartree-Fock calculations.  They find a purely electronic instability
for realistic yet narrow range of $U$ and $V$ that is outside the
values of $U$ and $V$ that they calculated using first-principles
constrained RPA calculations.  Watson \textit{et al.}\ have
demonstrated that the loss of the two mirror symmetries leads to band
hybridization and gap opening using polarization dependent
angle-resolved photoemission spectroscopy. They also performed density
functional theory (DFT) calculations that showed that the $B_{2g}$
atomic displacements found by Nakano \textit{et al.}\ reasonably
reproduce the electronic spectrum measured by them, and this leads
them to proposed that a structural instability, which could be either
due to an unstable elastic or zone-center optical phonon mode, causes the
phase transition in \tns.

The proposed electronic and structural instabilities should lead to
distinct temperature-dependent signatures in the measurement of the
zone-center phonon modes.  If the dynamical instability is due to an
unstable optical phonon mode, one would observe a softening of a
$B_{2g}$ phonon mode as the $T_c$ is approached from above, while no
such softening should be observed if the instability is electronic or
due to an unstable elastic mode corresponding to a uniform shear
distortion of the lattice.  A $B_{2g}$ optical phonon instability has
neither been observed in previous experiments nor been reported in
existing theoretical studies, and this motivates further
investigations of the dynamical properties of \tns.

In this paper, I present the results of first-principles calculation
of the phonon disperions of \tns\ performed using an
exchange-correlation functional that incorporates the van der Waals
interactions.  The dispersions exhibit two unstable optical branches
along the $\Gamma$--$Z$--$T$--$Y$--$\Gamma$ path, with the largest
instability occurring at $\Gamma$.  All the acoustic branches are
stable, which suggests that an elastic mode instability is not the
primary order parameter. The unstable modes have the irreps $B_{1g}$
and $B_{2g}$ at the zone center, with the $B_{1g}$ mode having a
larger magnitude of imaginary frequency.  However, atomic
displacements along the $B_{2g}$ mode results in larger spontaneous
strains than the displacements along the $B_{1g}$ mode.  After full
structural relaxations that minimize both the forces and stresses, the
$C2/c$ monoclinic structure due to the $B_{2g}$ phonon instability has
a lower total energy than the monoclinic structure that is stabilized
due to the $B_{1g}$ instability.  The theoretically obtained
structural parameters of the monoclinic phase agree well with the
experimentally determined ones.  Electronic structure calculations on
the monoclinic structure show a band gap that is comparable to the
values reported from experimental studies.  This suggests that the
unstable \btg zone-center optical phonon causes the phase transition
found in \tns, which should lead to an observation of a softening of
the \btg phonon mode as the $T_c$ is approached from above.

\section{Computational methods}

The structural relaxations and phonon calculations were performed using
the pseudopotential-based planewave method as implemented in the
Quantum {\sc espresso} software package \cite{qe}.  I used the
pseudopotentials generated by Dal Corso \cite{dalc14} and cutoffs of
60 and 600 Ry for the basis-set and charge-density expansions,
respectively.  A $12 \times 12 \times 4$ grid was used in the
Brillouin zone integration with a Marzari-Vanderbilt smearing of 0.01
Ry.  The dynamical matrices were calculated on an $8 \times 8 \times
4$ grid and Fourier interpolation was used to obtain the phonon
dispersions.  Most of the calculations presented here were performed
using the optB88-vdW exchange-correlation functional that accurately
treats the van der Waals interaction \cite{optb88}.  Some structural
relaxations were also performed using the LDA, PBE \cite{pbe}, and
PBEsol \cite{pbesol} functionals. Spin-orbit interaction was neglected
in these calculations.  I made extensive use of the {\sc amplimodes}
\cite{ampli}, {\sc spglib} \cite{spglib}, and {\sc phonopy}
\cite{phonopy} packages during the analysis of my results.

The electronic structure calculations were done using the {\sc elk}
software package \cite{elk}, which implements the general
full-potential linearized augmented planewave method.  I used the
Tran-Blaha modified Becke-Johnson (mBJ) potential in these
calculations to obtain improved band gaps \cite{mbj}. A $12 \times 12
\times 4$ grid was used in the Brillouin zone integration with a
Fermi-Dirac smearing of 0.002 Ry.  The cutoffs for the planewaves in
the basis-set and charge-density expansions were determined by the
parameters $RK_{\textrm{max}} = 8$ and $G_{\textrm{max}} = 16$,
respectively. Spin-orbit interaction was included in these
calculations.

\section{Results and discussion}

\begin{figure}
  \includegraphics[width=0.8\columnwidth]{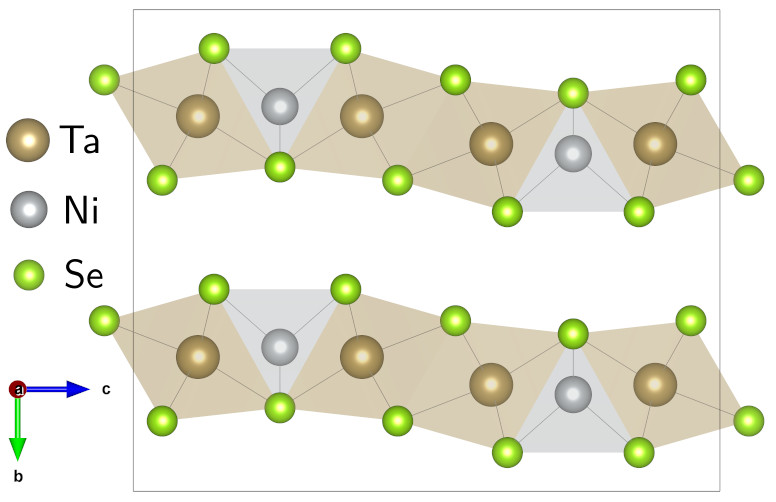}
  \includegraphics[width=0.8\columnwidth]{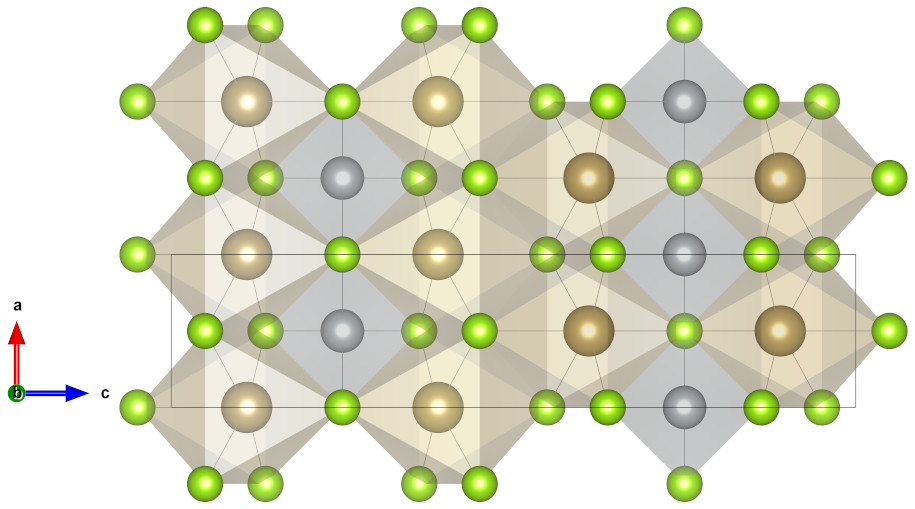}
  \caption{Crystal structure of orthorhombic \tns\ viewed along two axes.}
  \label{fig:orth-struct}
\end{figure}

\tns\ has a layered structure \cite{suns85,disa86}, which is shown in
Fig.~\ref{fig:orth-struct}.  Each layer is three atoms thick and is
composed of Se sheets at the top and bottom sandwiching a Ta/Ni sheet.
The Se atoms octahedrally and tetrahedrally coordinate the Ta and Ni
atoms, respectively.  These polyhedra are stacked such that the Ta and
Ni atoms form chains along the $a$ axis.  Along the $c$ axis, each Ni
chain is separated by two Ta chains.  This suppresses hopping along
the $c$ axis.  Thus, the electronic structure exhibits a
quasi-one-dimensional feature with dispersions that are smaller along
the $b$ and $c$ axes than along the $a$ axis that is parallel to the
chain direction \cite{kane13,lark17}.

\begin{table}
  \caption{\label{tab:latt} Calculated lattice parameters of
    orthorhombic \tns\ obtained using different exchange-correlation
    functionals. The volume is given per formula unit.}
  \begin{ruledtabular}
     \begin{tabular}{l d{1.3} d{1.3} d{1.3} d{1.3}}
       functional & \multicolumn{1}{c}{$a$ (\AA)} & \multicolumn{1}{c}{$b$ (\AA)}
       & \multicolumn{1}{c}{$c$ (\AA)} & \multicolumn{1}{c}{volume
         (\AA$^3$/f.u.)} \\
       \hline
       LDA        & 3.4021 & 12.6548 & 15.3822 & 331.12 \\
       PBE        & 3.5040 & 14.2895 & 15.7444 & 394.17 \\
       PBEsol     & 3.4325 & 13.0700 & 15.4981 & 347.65 \\
       optB88-vdW & 3.5075 & 13.0352 & 15.7537 & 360.14 \\
       Experiment\footnote{Ref.~\cite{naka18b}.} & 3.5029 & 12.8699 &
       15.6768  & 353.37
    \end{tabular}
  \end{ruledtabular}
\end{table}

\begin{table}
  \caption{\label{tab:oint} Calculated internal atomic parameters of
    orthorhombic \tns\ obtained using the opt88B-vdW functional.}
  \begin{ruledtabular}
    \begin{tabular}{l l @{\hspace{1em}} d{1.1} d{1.5} d{1.5} @{\hspace{1em}} d{1.1} d{1.6} d{1.6}}
      & &  \multicolumn{3}{c}{theory}  & \multicolumn{3}{c}{experiment\footnote{Ref.~\cite{naka18b}}} \\
      atom & site  & \multicolumn{1}{c}{$x$} & \multicolumn{1}{c}{$y$}
       & \multicolumn{1}{c}{$z$} & \multicolumn{1}{c}{$x$} & \multicolumn{1}{c}{$y$}
       & \multicolumn{1}{c}{$z$} \\
       \hline
       Ta    & $8f$ & 0.5 & 0.2216 & 0.1114  & 0.5 & 0.221158 & 0.110222 \\
       Ni    & $4c$ & 0.0 & 0.2029 & 0.25    & 0.0 & 0.20096  & 0.25 \\
       Se(1) & $8f$ & 0.5 & 0.3285 & 0.25    & 0.5 & 0.32679  & 0.25 \\
       Se(2) & $8f$ & 0.0 & 0.3532 & 0.0486  & 0.0 & 0.354170 & 0.049338 \\
       Se(3) & $4c$ & 0.0 & 0.0814 & 0.1386  & 0.0 & 0.080461 & 0.137726 \\
    \end{tabular}
  \end{ruledtabular}
\end{table}

The calculated lattice parameters of orthorhombic \tns\ obtained from
full structural relaxation calculations using various
exchange-correlation functionals along with the experimental ones
\cite{naka18b} are shown in Table~\ref{tab:latt}.  The LDA
underestimates the volume by 6.3\%, while the PBE overestimates it by
11.6\%.  The PBEsol yields a volume that is closer to the experimental
value.  It underestimates the volume by 1.6\%, but it also gives a
larger out-of-plane lattice constant $b$.  The optB88-vdW functional
overestimates the volume by 1.9\%, which is a slightly worse
performance compared to the PBEsol.  However, the optB88-vdW lattice
parameters are uniformly closer to the experimental values compared to
the PBEsol lattice parameters.  It is interesting to note that the PBE
gives values for the in-plane lattice constants $a$ and $b$ that are
closest to the experiment despite overestimating the out-of-plane
lattice constant $b$ by 11.0\%.  The optB88-vdW improves the
description of the van der Waals interaction and gives a value of $b$
that is much closer to the experiment.  The calculated internal atomic
parameters obtained using this functional and the experimentally
determined ones \cite{naka18b} are given in Table~\ref{tab:oint}, and
the respective values are in good agreement with each other.  Since
optB88-vdW is structurally the most accurate functional among the four
that I tested, further investigations of the dynamical properties of
\tns\ are performed using it.

\begin{figure}
  \includegraphics[width=0.7\columnwidth]{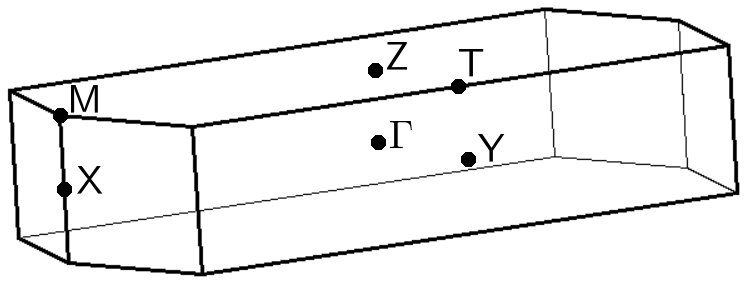}
  \includegraphics[width=\columnwidth]{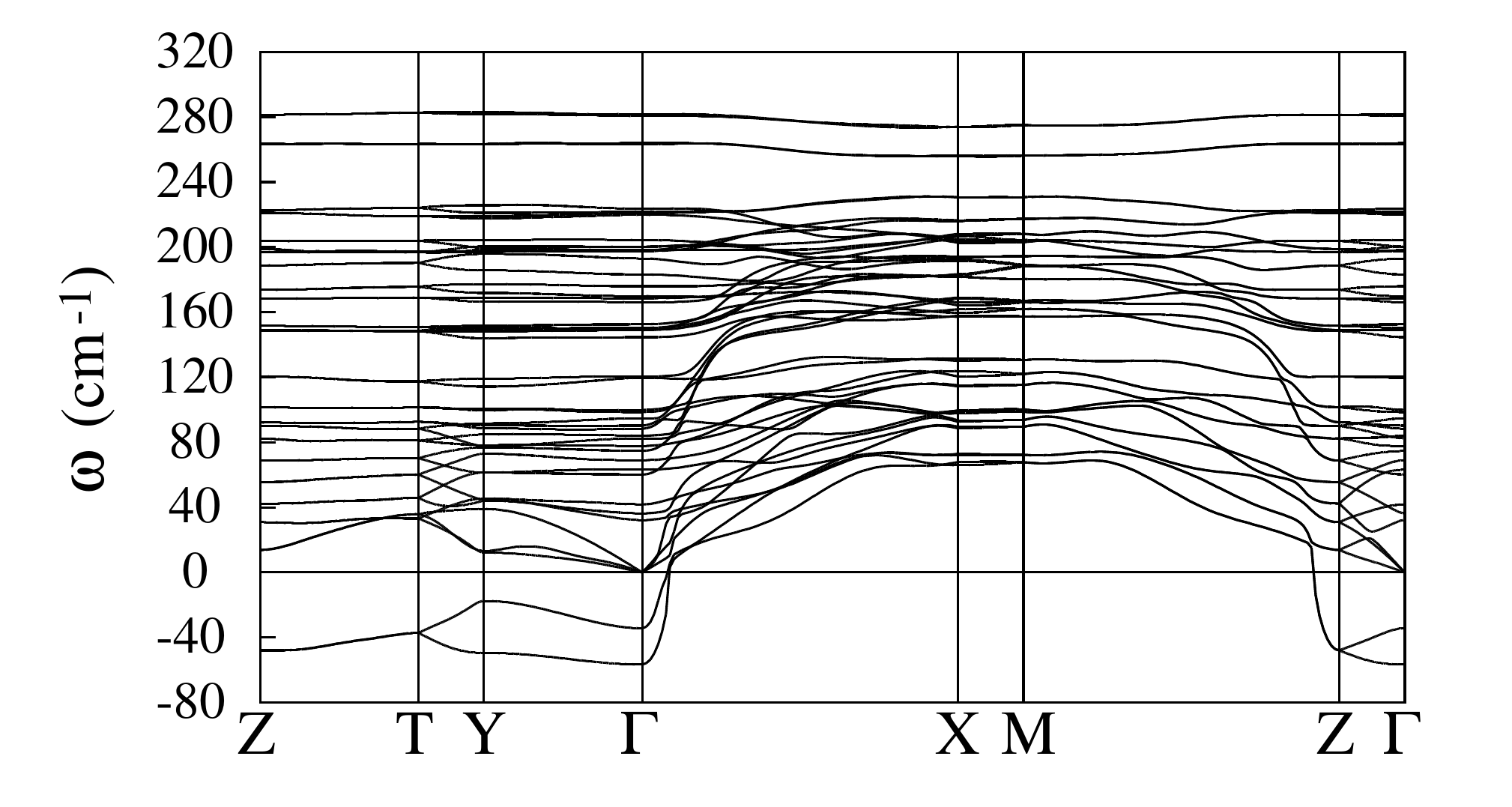}
  \caption{(Top) The Brillouin zone of orthorhombic \tns.  (Bottom)
    Calculated phonon dispersions of fully-relaxed orthorhombic \tns\ 
    obtained using the optB88-vdW functional.}
  \label{fig:orth-ph}
\end{figure}

Fig.~\ref{fig:orth-ph} shows the calculated phonon dispersions of the
fully-relaxed orthorhombic \tns, as well as its Brillouin zone
depicting the high-symmetry points.  The dispersions show two unstable
optical branches along the $\Gamma$--$Z$--$T$--$Y$--$\Gamma$ path.
All the acoustic branches are stable, which suggests that the
structural transition observed in this material is not primarily due
to an elastic instability.  The largest optical phonon instability is
at $\Gamma$, in agreement with the experimental finding that the
unit-cell size does not multiply across the transition
\cite{suns85,disa86}.  The most unstable mode at $\Gamma$ has the
irrep \bog and a calculated frequency of 56$i$ cm$^{-1}$. The other
unstable mode is of \btg irrep and has a frequency of 34$i$ cm$^{-1}$.



The \bog and \btg optical phonon instabilities lead to $P2_1/m$ and
$C2/c$ structures, respectively. This result showing the \bog mode to
have a larger instability than the \btg mode is at odds with the
experiments that find the low-temperature structure to be $C2/c$
\cite{suns85,disa86,naka18a}.  I displaced the atomic positions of the
orthorhombic $Cmcm$ structure according to the eigenvectors of the
\bog and \btg modes and performed structural relaxation calculations
within the optB88-vdW functional to see which low-symmetry structure
is more energetically stable.  When only the atomic positions are
relaxed by minimizing the forces, the $P2_1/m$ structure is more
stable than the $C2/c$ structure.  However, the relaxation of only the
atomic positions results in the development of stresses, which are
larger in the $C2/c$ structure.  When full structural relaxation is
performed by minimizing both the forces and stresses, I find that the
$C2/c$ structure is more stable than the $P2_1/m$ structure.  The
calculated energy difference between the two structures is small, with
a value of 26.0 $\mu$eV per formula unit.  The energies of these
monoclinic structures are $\sim$1 meV per formula unit lower than that
of the orthorhombic structure.

\begin{figure}
  \includegraphics[width=0.95\columnwidth]{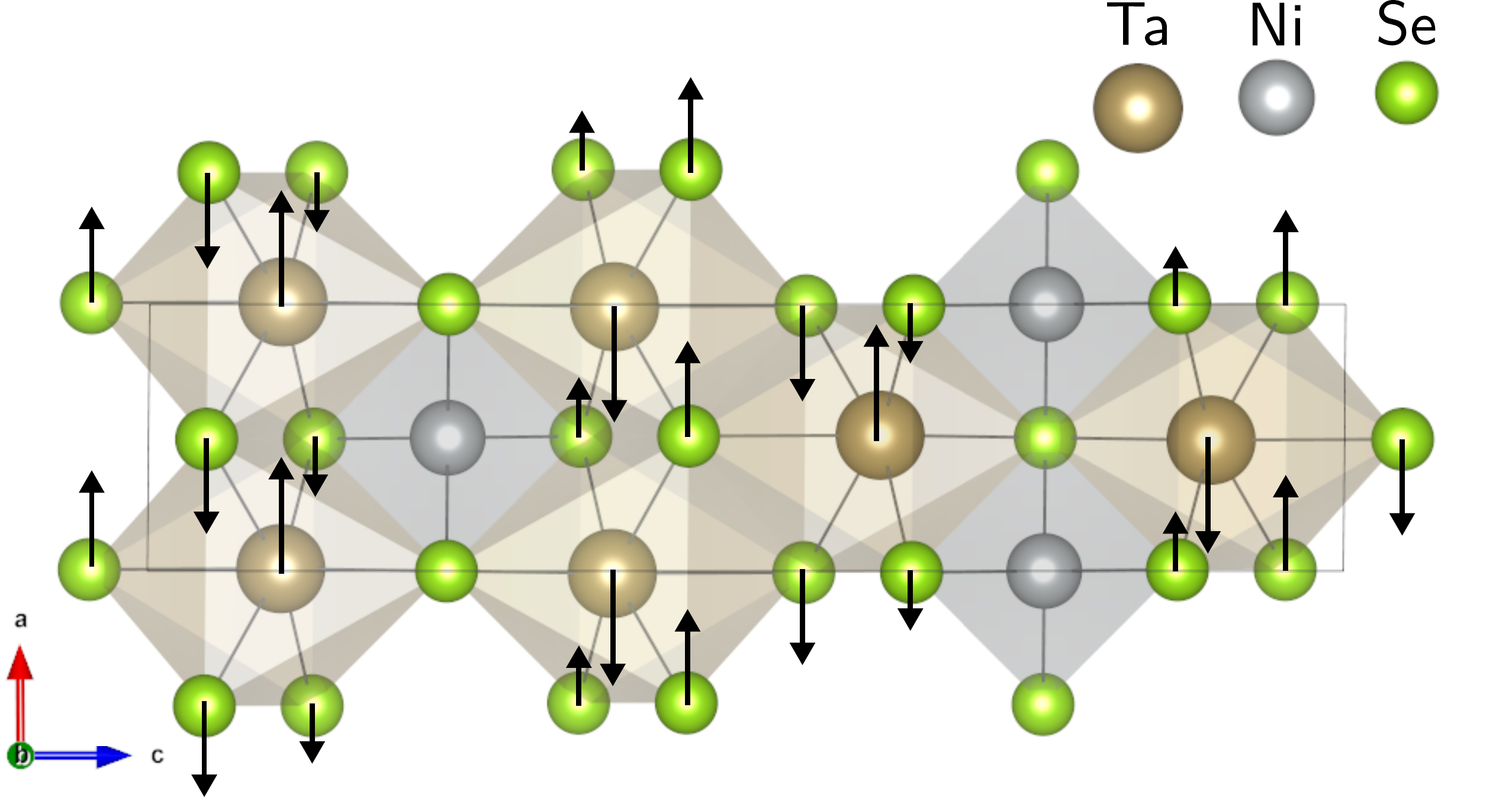}
  \caption{Displacement pattern of the unstable $B_{2g}$ optical
    phonon mode of orthorhombic \tns.}
  \label{fig:orth-ph-disp}
\end{figure}

\begin{table}
  \caption{\label{tab:mono} Calculated internal atomic parameters of
    monoclinic \tns\ obtained using the opt88B-vdW
    functional. Calculated lattice parameters are
    $a$ = 3.510, $b$ = 13.037, $c$ =  15.754 \AA, $\beta =
    90.492^\circ$.  The experimental lattice parameters are
    $a$ = 3.4916, $b$ = 12.814, $c$ =  15.649 \AA, $\beta =
    90.693^\circ$\footnote{Ref.~\cite{naka18a}.}. 
    }
  \begin{ruledtabular}
    \begin{tabular}{l l @{\hspace{1ex}} d{1.4} d{1.4} d{1.4} @{\hspace{1ex}} d{1.5} d{1.6} d{1.6}}
      & &  \multicolumn{3}{c}{theory}  &
      \multicolumn{3}{c}{experiment} \\
      atom & site  & \multicolumn{1}{c}{$x$} & \multicolumn{1}{c}{$y$}
       & \multicolumn{1}{c}{$z$} & \multicolumn{1}{c}{$x$} & \multicolumn{1}{c}{$y$}
       & \multicolumn{1}{c}{$z$} \\
       \hline
Ta    & $8f$ & 0.4909  & 0.2218 & 0.6114   &  0.48795 & 0.221366 & 0.610655  \\
Ni    & $4e$ & 0.0     & 0.2031 & 0.75     &  0.0     & 0.201210 & 0.75     \\
Se(1) & $8f$ & 0.0058  & 0.3533 & 0.5486   &  0.00861 & 0.354601 & 0.548924  \\
Se(2) & $8f$ & 0.0052  & 0.0817 & 0.6387   &  0.00825 & 0.079931 & 0.638047  \\
Se(3) & $4e$ & 0.5     & 0.3287 & 0.75     &  0.5     & 0.327557 & 0.75     \\

    \end{tabular}
  \end{ruledtabular}
\end{table}

The atomic displacements obtained from the eigenvector of the unstable
\btg optical phonon mode are shown in Fig.~\ref{fig:orth-ph-disp}, and
the fully-relaxed structural parameters of the $C2/c$ structure are
given in Table \ref{tab:mono}.  These are in agreement with the
distortions in the low-temperature $C2/c$ structure reported by Nakano
\textit{et al.}\ using synchtron x-ray diffraction experiments
\cite{naka18a}.  They suggested that atomic displacements in the
$C2/c$ structure result from the condensation of two \btg optical
phonon modes.  In contrast, I find that only one \btg optical phonon
instability is responsible for the transition to the low-temperature
structure.  This \btg mode causes the Se ions at two $8f$ Wyckoff
sites to displace out-of-phase with the Ta ions at the $8f$ site along
the chain direction $a$ axis and leads to the loss of both the $m_x$
and $m_z$ mirror symmetries, whereas the Ni and Se ions at the $4c$
Wyckoff sites do not move.  This displacement pattern results in the
formation of an electric dipole moment in the TaSe$_6$ octahedra.
However, the phase of the atomic displacements is opposite in the
TaSe$_6$ octahedral chains that lie on either side of the NiSe$_4$
tetrahedra.  As pointed out by Nakano \textit{et al.}  \cite{naka18a},
this results in an intra-unit-cell antiferroelectric distortion that
preserves the global inversion symmetry in the low-temperature phase
of \tns.  In addition to the Ta and Se ions moving against each other,
the $B_{2g}$ mode causes the two different sets of Se ions at $8f$
positions to move by different amounts.  This causes a slight shear
distortion of the TaSe$_6$ octahedra, and the phase of the shear
distortion is opposite for the nearest-neighbor octahedra along the
$c$ axis.

\begin{figure}
  \includegraphics[width=\columnwidth]{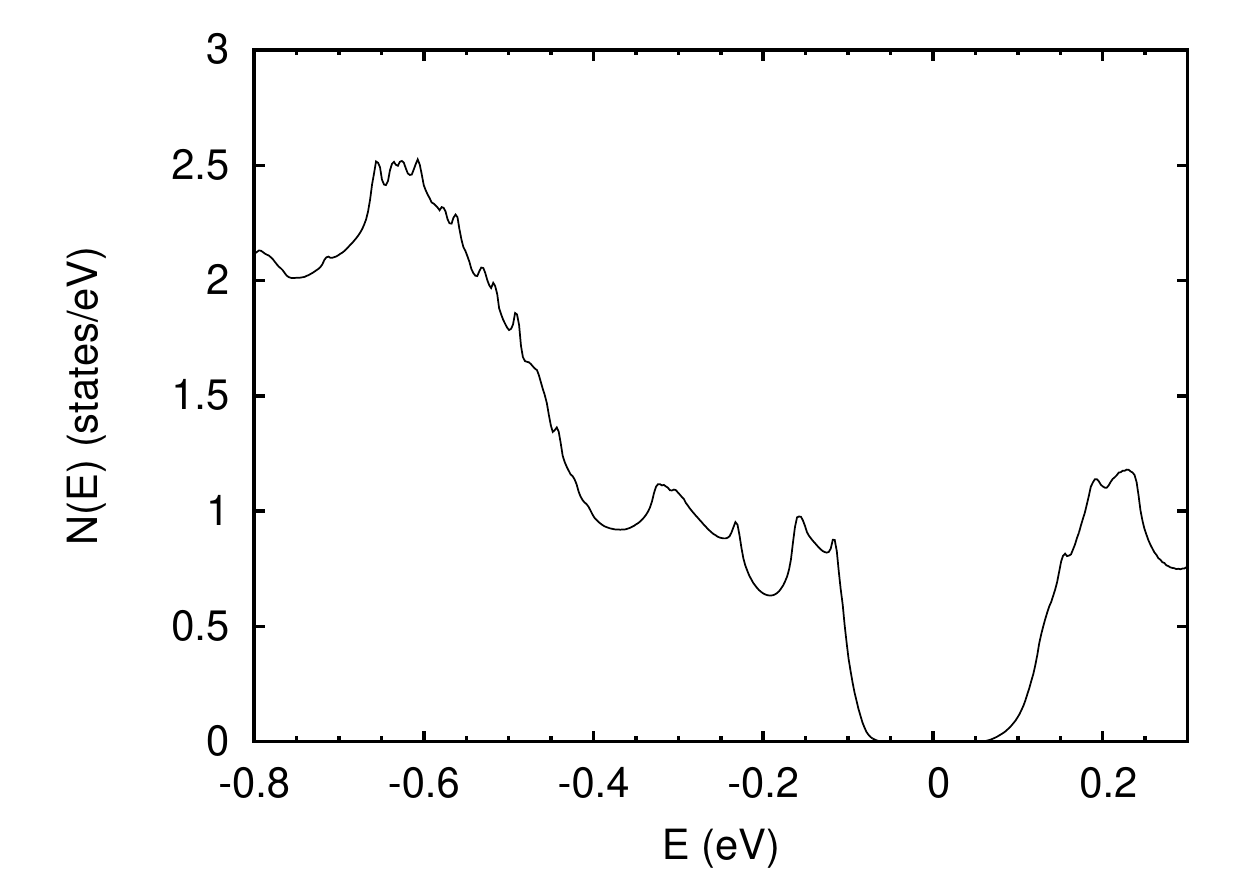}

  \includegraphics[width=\columnwidth]{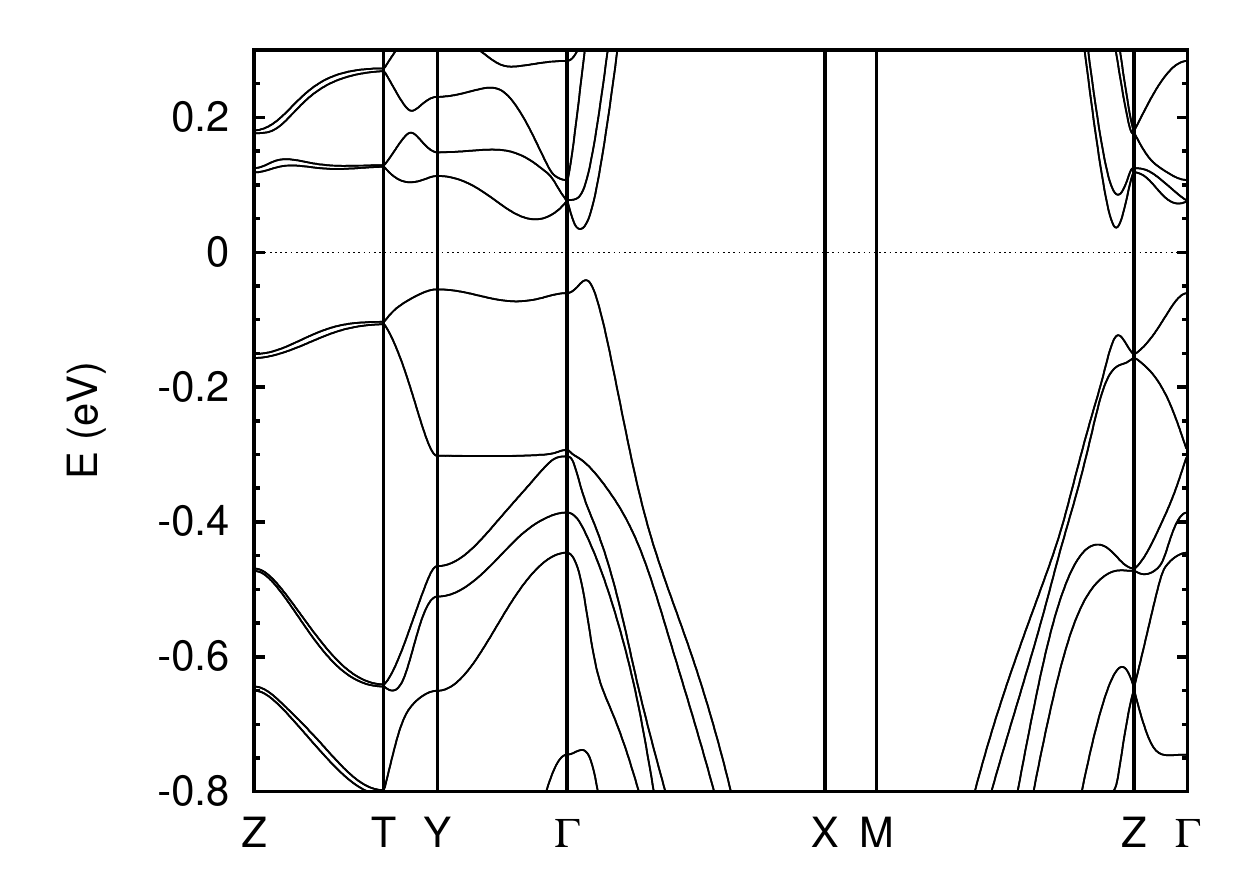}
  \caption{(Top) Electronic density of states per forumla unit and
    (bottom) band structure of fully relaxed monoclinic
    \tns\ calculated using the mBJ potential.}
  \label{fig:mband}
\end{figure}

The electronic density of states and band structure of monoclinic
$C2/c$ \tns\ calculated using the mBJ potential are shown in
Fig.~\ref{fig:mband}.  While the optB88-vdW functional is constructed
to accurately yield the structural parameters of layered materials,
the mBJ potential has been developed to give improved band gaps.  The
mBJ band structure reported here reasonably agrees with the one
calculated by Watson \textit{et al.}  \cite{wats19}, although there
are some qualitative and quantitative differences. I obtain a gap at
$\Gamma$ of 137 meV.  The value of the smallest direct and indirect
gaps are 80 and 74 meV, respectively, both of which occur near
$\Gamma$ along the $\Gamma$--$X$ path.  In the density of states, the
first peaks below and above the gap are at the relative values of
$-$117 and 192 meV, respectively, with a gap of 309 meV between the
peaks. For comparison, a gap of 160 meV is observed for optical
excitations in spectroscopy experiments \cite{lu17,lark17} and 300 meV
is found in scanning tunneling spectroscopy experiment \cite{lee19}.

Neither the calculated valence nor conduction band is flat near
$\Gamma$ along the $\Gamma$--$X$ path, and their respective band edges
lie slightly away from $\Gamma$.  A curvature in the valence band edge
near $\Gamma$ has been observed by Watson \textit{et al.}\ in their
ARPES experiment \cite{wats19}, and this contradicts the expectation
of a flat band in an excitonic condensate phase. Since DFT
calculations show a sizeable gap opening in the presence of $B_{2g}$
shear and atomic displacements, Watson \textit{et al.}\ have argued
that the phase transition in \tns\ is due to a structural instability.
The advancement made in the present paper is the identification of the
\btg zone-center optical phonon mode depicted in
Fig.~\ref{fig:orth-ph-disp} as the primary order parameter for the
transition to the low-temperature phase.  This \btg dynamical
instability is also in accord with the presence of a monoclinic
distortion below $T_c$ that arises due to the softening of the
$c_{55}$ elastic modulus.  The corresponding elastic strain
$\varepsilon_{xz}$ and the unstable phonon mode have the same \btg
irrep, and this allows a linear coupling between the two.  When there
is such a term in the free energy, Landau theory analysis shows that
the elastic modulus also diverges as the $T_c$ is approached
\cite{rehw73}, resulting in the coexistence of both the \btg
distortions.  Nakano \textit{et al.}\ have observed a softening of the
transverse acoustic mode corresponding to the $c_{55}$ elastic modulus
near transition \cite{naka18a}, but they did not see a softening of
the only \btg optical phonon mode that they could resolve between 30
and 100 cm$^{-1}$ in the high-temperature orthorhombic phase.  There
are three \btg optical phonon modes corresponding to three
chain-direction out-of-phase displacements of the three $8f$ Wyckoff
sites present in the structure.  If the zone-center optical phonon
mode is the primary instability, one should observe a softening of the
\btg optical mode shown in Fig.~\ref{fig:orth-ph-disp} before the
acoustic mode starts softening when the $T_c$ is approached from high
temperatures.

The calculated frequencies of all zone-center optical phonon modes of
the orthorhombic and monoclinic phases are given in
Table \ref{tab:freq} in Appendix.  In the orthorhombic phase, the three \btg modes
have frequencies of $34i$, 75 and 149 cm$^{-1}$.  The related
modes in the monoclinic phase have the irrep $A_g$.  By comparing the
eigenvectors of the phonons in the two phases, I find the
correspondence $34i \rightarrow 99$ cm$^{-1}$, $75
\rightarrow 60$ cm$^{-1}$, and $149 \rightarrow 145$
cm$^{-1}$ between the \btg phonons in the orthorhombic phase and the
$A_g$ phonons in the monoclinic phase.  Nakano \textit{et
  al.}\ measured phonons up to 120 cm$^{-1}$ and observed only the
75 cm$^{-1}$ \btg mode in the orthorhombic phase between 400 and
$T_c = 328$ K.  They did not observe the unstable \btg mode presumably
because this mode is already too soft and overdamped below 400 K.
Below $T_c$ they observe an new mode at 88 cm$^{-1}$.  Yan \textit{et
  al.}\ also report the appearance of an extra $A_g$ phonon below
$T_c$ in their Raman scattering study \cite{yan19}.  I claim that this
new mode in the monoclinic phase corresponds to the amplitude
modulation of the order parameter deriving from the unstable \btg mode
of the orthorhombic phase.

DFT calculations approximately incorporate the many-body exchange and
correlation interactions arising out of the Coulomb repulsion between
electrons, but they do not describe the many-body phenomena of
electron-hole pairing and pair condensation that is associated with an
excitonic insulator.  Therefore, the dynamical instability described
here is different from the excitonic instability proposed by Wakisaka
\textit{et al.}\ based on their observation of a flat band in the
low-temperature phase \cite{waki09}.  A purely electronic excitonic
mechanism for the phase transition in \tns\ has been supported by a
body of theoretical studies
\cite{kane13,ejim14,sugi16a,yama16,domo16,sugi16b,mats16,domo18,sugi18},
but such an electronic instability or, indeed, an optical phonon one
has not been unambiguously observed in any experimental study.  As
discussed above, a mechanism that leads to a complex order parameter
cannot be the cause of the phase transition because the transition
involves the loss of symmetry operations belonging to the $B_{2g}$
irrep.  Hence, among the theoretical studies that support a purely
electronic mechanism, the one by Mazza \textit{et al.}\ \cite{mazz19}
is remarkable because they not only use the hopping and interaction
parameters derived from first principles but also allow the breaking
of the $m_x$ and $m_z$ mirror symmetries that have been previously
reported by a synchroton x-ray diffraction experiment \cite{naka18a}.
They find that an electronic order parameter that breaks the mirror
symmetries has a finite value for a narrow range of Coulomb
interaction parameters $U$ and $V$.  Although the range is realistic,
the calculated values of $U$ and $V$ that they obtained from their
first-principles constrained RPA calculations lie outside this range.
Therefore, a reasonable interpretation of their result is that an
electronic mechanism does not underlie the phase transition observed
in \tns, which further gives credence to a mechanism based on a
dynamical instability.

Ultimately, the nature of the instability that drives the phase
transition in \tns\ can only be validated by experiments.  In this
regard, electronic and phonon instabilities will display different
responses in inelastic scattering experiments.  The presence of a \btg
zone-center optical phonon instability will be reflected by a
softening of this mode as the $T_c$ is approached from above.  This
optical mode should soften before the electronic and elastic modes
belonging to the \btg irrep also start becoming unstable near the
$T_c$. On the other hand, if the transition is due to either
electronic or elastic instability, none of the zone-center \btg
optical phonon modes should soften as the temperature is lowered.



\section{Summary and Conclusions}

This paper presents the calculated phonon dispersions of \tns\ in the
high-temperature orthorhombic $Cmcm$ phase.  They exhibit two optical
phonon branches that are unstable along the
$\Gamma$--$Z$--$T$--$Y$--$\Gamma$ path.  The largest instability is at
the zone center, consistent with the absence of an increase in the
size of the unit cell across the transition as deduced from x-ray
diffraction studies.  All the acoustic branches are stable, which
indicates that a zone-center optical instability is the primary
instability driving the transition.  The two unstable modes at the
zone center have the \bog and \btg irreps.  Full structural
relaxations minimizing both the forces and stress show that the
monoclinic $C2/c$ structure corresponding to the \btg instability is
lowest in energy, and this structure exhibits a sizable band gap.
Although my DFT calculations cannot compute purely electronic
instabilities, the fact that the ground state phase obtained in the
present study correctly describes the experimentally determined
structural and electronic properties of the low-temperature phase
suggests that a zone-center optical phonon instability is responsible
for the phase transition in \tns.  An observation of a \btg
zone-center optical phonon softening before the enhancement of \btg
electronic and elastic instabilities near $T_c$ would confirm the
mechanism proposed here.  However, if none of the \btg optical phonons
present in the material soften as the $T_c$ is approached from above,
this would imply that the transition is due to either electronic or
elastic instability.

\section{acknowledgements}

I am grateful to B. J. Kim and Indranil Paul for insightful
discussions.  This work was supported by GENCI-CINES (grant
2019-A0070911099), the Swiss National Supercomputing Center (grant
s820), and the European Research Council (grant ERC-319286 QMAC).

\appendix

\begin{table}[b]
  \caption{\label{tab:freq} Calculated frequencies of zone-center
    optical phonon modes of orthorhombic and monoclinic \tns.}
  \begin{ruledtabular}
    \begin{tabular}{ll@{\qquad}ll}
      \multicolumn{2}{c}{orthorhombic} &
      \multicolumn{2}{c}{monoclinic} \\
      irrep & frequency (cm$^{-1}$) & irrep & frequency (cm$^{-1}$) \\
      \hline
      \bog   &     56.537$i$  &    \ag    &       32.448    \\  
      \btg   &     34.282$i$  &    \bu    &       37.300    \\
      \ag    &     32.272     &    \bg    &       37.954    \\
      \bou   &     36.411     &    \au    &       41.872    \\
      \au    &     41.880     &    \ag    &       59.704    \\
      \bog   &     60.292     &    \bg    &       62.443    \\
      \brg   &     63.250     &    \bg    &       70.627    \\
      \brg   &     69.073     &    \bu    &       79.896    \\
      \btg   &     74.755     &    \bg    &       81.898    \\
      \bru   &     77.810     &    \au    &       82.490    \\
      \btu   &     82.608     &    \bg    &       91.151    \\
      \bog   &     84.121     &    \bu    &       93.610    \\
      \au    &     88.425     &    \ag    &       95.020    \\
      \bru   &     90.366     &    \ag    &       99.424    \\
      \ag    &     94.880     &    \bg    &      101.732    \\
      \brg   &     98.708     &    \au    &      103.982    \\
      \bou   &    100.029     &    \bu    &      116.820    \\
      \btu   &    119.832     &    \au    &      119.502    \\
      \ag    &    120.515     &    \ag    &      120.569    \\
      \bou   &    144.638     &    \bu    &      141.771    \\
      \au    &    149.112     &    \au    &      144.654    \\
      \btg   &    149.189     &    \ag    &      145.162    \\
      \btu   &    149.228     &    \bg    &      149.436    \\
      \bru   &    150.490     &    \au    &      152.897    \\
      \bog   &    152.653     &    \bu    &      154.043    \\
      \bou   &    166.164     &    \bu    &      167.925    \\
      \brg   &    168.567     &    \bg    &      168.780    \\
      \ag    &    169.927     &    \ag    &      170.196    \\
      \brg   &    176.025     &    \bg    &      176.088    \\
      \ag    &    183.266     &    \ag    &      184.067    \\
      \btu   &    193.170     &    \au    &      194.021    \\
      \bru   &    197.661     &    \bg    &      197.578    \\
      \bog   &    198.033     &    \bu    &      197.609    \\
      \bou   &    198.304     &    \bu    &      199.532    \\
      \brg   &    200.367     &    \bg    &      203.609    \\
      \ag    &    200.463     &    \ag    &      203.635    \\
      \btu   &    204.457     &    \au    &      206.125    \\
      \brg   &    220.011     &    \bg    &      222.413    \\
      \ag    &    220.940     &    \ag    &      222.760    \\
      \btu   &    221.746     &    \au    &      223.155    \\
      \bou   &    223.661     &    \bu    &      226.275    \\
      \bou   &    263.460     &    \bu    &      264.929    \\
      \brg   &    264.017     &    \bg    &      266.159    \\
      \ag    &    280.743     &    \ag    &      281.137    \\
      \btu   &    282.049     &    \au    &      282.199   
     \end{tabular}
  \end{ruledtabular}
\end{table}

\end{document}